\documentclass[12pt,oneside]{article}

\usepackage{amsmath}
\usepackage{amsfonts}
\usepackage[utf8]{inputenc}
\usepackage{mathtools}
\usepackage{amssymb}
\usepackage[pdftex]{graphicx}
\usepackage{epstopdf}
\usepackage{subcaption}
\usepackage[numbers]{natbib}
\usepackage{fancyhdr}
\usepackage[english]{babel}
\usepackage{url}
\graphicspath{ {Img/} }
\usepackage{setspace}
\usepackage[breaklinks=true, bookmarks=true, bookmarksopen=true, colorlinks=true, linkcolor=blue, citecolor=red, pdfhighlight=/P]{hyperref}
\usepackage[a4paper,top=1in, bottom=1in, left=1in, right=1in]{geometry}
\usepackage{makeidx}
\usepackage[inlinechap]{fncystyle}
\usepackage{float}
\usepackage{braket}
\usepackage{physics}
\usepackage{bm}
\usepackage{bbm}
\usepackage{bbold}
\usepackage{tikz}
\usetikzlibrary{shapes, arrows}
\usepackage{enumitem}
\newlist{numeration}{enumerate}{10}
\setlist[numeration]{label*=\arabic*.}
\usepackage{amssymb}
\usepackage{caption}
\usepackage[utf8]{inputenc}
\usepackage[T1]{fontenc}
\usepackage{authblk}

\tikzstyle{block} = [rectangle, draw, fill=blue!20, text width=10em, text centered, rounded corners, minimum height=4em]
\tikzstyle{line} = [draw, -latex']

\usepackage{stackengine} %para wavy conjugate

\newcommand{\s}{\sigma}

\newcommand{\da}{\dagger}
\newcommand{\down}{\downarrow}
\newcommand{\up}{\uparrow}
\newcommand{\bb}[1]{ \boldsymbol{#1}}

\newcommand{\sla}[1]{\not\!{#1}}
\newcommand{\I}{\bb{i}}
\newcommand{\J}{\bb{j}}
\newcommand{\K}{\bb{k}}
\newcommand{\II}{{\text{\textbf{i}}}}
\newcommand{\JJ}{{\text{\textbf{j}}}}
\newcommand{\KK}{{\text{\textbf{k}}}}

\newcommand{\E}{\bb{e}}
\newcommand{\bo}{\bb{o}}
\newcommand{\g}{\bb{\gamma}\!\!\!\!\bb{\gamma}}
\newcommand{\Eq}[1]{ E.q(\ref{#1})}

\usepackage{accents}
\newcommand{\thickbar}[1]{\accentset{\rule{.4em}{1.2pt}}{#1}}

\title{ A unified field theory from a complexified quaternion-octonion Dirac equation}
\date{April 2022}

\author[1]{Juan Camilo V\'elez Qui\~{n}ones \thanks{juan.camilo.velez@correounivalle.edu.co}}
\affil[1]{Physics Department, Universidad del Valle, Cali, Colombia.}

\begin{document}

\maketitle

\begin{abstract}
	 It is set manifest an underlying algebraic structure of Dirac equation and solutions, in terms of C$\ell_2$ Clifford algebra projectors and ladder operators. From it, a scheme is proposed for constructing unified field theories by enlarging the pointed algebra. A toy unified matter field model is formulated, modifying the Dirac equation with complex quaternions and octonions. The result describes a set of fermion fields with reminiscent properties of one standard model particle generation, exhibiting U(1) electromagnetism, SU(2) flavor, and  SU(3) color symmetries, remarkably SU(2) induces frame fields, though further explorations are needed. 
\end{abstract}

\section*{ \center I. INTRODUCTION}

In particle physics, unified field theories (UFTs) try to describe fundamental particles and forces of nature as different manifestations of a single phenomenon aiming to elucidate underlying principles, behind seeming complexity. Throughout the last decades, these theories have been applied as an alternative for tackling the apparent structure arbitrariness of the standard model, by merging elementary particles and interactions through simple principles that could explain its rather ad hoc appearance \cite{boer1994,croon2019}. 

The roots of the unification pursuit lie in the successful unification of electromagnetism and weak force into electroweak theory in the 60s. From there the natural question was raised, if it is possible to encompass strong force together with electroweak force into a grand unified theory (GUT). With the aforementioned motivation, further attempts have been driven for constructing theories meant to merge the standard model interactions into a single force.

The first GUTs were the four-color Pati Salam model \cite{pati1974}, Georgi and Glashow SU(5) \cite{georgi1974} fusing all interactions in an electrocolor force, and the Georgi \cite{georgi1975}/Fritzsch and Minkowski \cite{minkowski1975} Spin(10), which go further not only unifying bosons but fermions as well. These models give account in general for one generation of standard model particles and their interactions. This comes at a cost, like the intrusion of new interaction and proton decay predictions, which remain experimentally unobserved up today. Despite these issues, a plethora of theories have been formulated since then, but the problems from initial works persist.

The usual approach for formulating GUTs is an extension of the methodology used for the elaboration of electroweak theory: a master gauge group containing U(1), SU(2), and SU(3) is proposed, and from such a bigger group, an invariant Lagrangian under gauge transformations is crafted, requiring the insertion of particle fields and interactions.

However, the mentioned GUT's methodology is not the only one for conceiving unified theories. An alternative is an algebraic approach, taking as a starting point algebras, instead of master gauge groups. There, particles correspond to algebra elements where inherent symmetries and hence charges, emerge naturally in relation to actions of other algebra elements. This can be done by elaborating minimal left ideals for representing particles \cite{chisholm1999, trayling1999, Furey2014,Furey2015, Furey2016, stoica2018}, but it is not strictly necessary \cite{trayling2001}. The referred constructions can describe different forces emerging from algebra symmetries, without appealing to a master group as GUTs do, therefore avoiding a source of unknown forces and unwanted effects \cite{trayling1999}. Nevertheless, they could result in an inaccurate description, not including all forces, or not getting standard model symmetry groups. 

Early examples of this algebraic approach were for instance the strong interaction descriptions using octonions \cite{Giirsey1974, katsusada1981,penney1971}, later extended to other forces using the four normed division algebras, which are the real $\mathbb{R}$, complex $\mathbb{C}$,  quaternions $\mathbb{H}$ and octonions $\mathbb{O}$ numbers \cite{dixon1994}. 

In the algebra approach, the usage of Clifford algebras is common. Among their advantages is the fact that spinors are left or right ideals \cite{Porteous1995}, making them a suitable language for describing fermions, and for esthetic equations, which can give algebraic structure inside \cite{barducci1977, Gendve1979, casalbuoni1980}. Additionally, division algebras are related to them through isomorphisms between quaternions, octonions, and C$\ell_2$, C$\ell_6$ Clifford algebras respectively \cite{sultan2013, Furey2016}. Early examples of implementations of these algebras are the theories of Chisholm and Farwell \cite{chisholm1989}, or more updated models include C$\ell_6$ Stoica's \cite{stoica2018}, or Furey's \cite{Furey2015, Furey2016} casted with octonions, but isomorphic to C$\ell_6$. 

The algebra approach it is sometimes not addressed the why and how the algebraic nature of particles could interplay in Lagrangians, ending with a lack of dynamical equations. However when it is tackled, it could yield that Lagrangians allow being casted in modified versions, where the Dirac gamma matrices are replaced by elements belonging to the algebra \cite{trayling1999, trayling2001, chisholm1989, chisholm1999, penney1971}, resulting in single expressions, which contain all fermion fields from one generation. The fact that altered Lagrangians could describe several types of particles in a single field \cite{penney1971,trayling2001,chisholm1999,giardino2021,stefano1996}, is not unexpected since enlarging gamma matrices imply a wider dimension for the Dirac equation solutions, a desirable feature for fermions unification.

Relative to what has been said, here is an exploration of an alternative course of action for generating a UFT via modifying the Dirac equation (DE). This is achieved through a concrete example where DE is reformulated with complexified quaternions and octonions, resulting in a theory with a Dirac-like equation, that simultaneously describes all fermions from one generation with a unified matter field, and forces are included by Yang-Mills covariant derivatives. All of these heavily rely on Furey's \cite{Furey2015, Furey2016} previous results. The procedure is generalized in a straightforward manner, being an alternative scheme to GUTs master gauge groups, and serving as a complement to the algebraic approach, providing a systematic method of dynamical equations formulation for any algebra with a Clifford sub-algebra.

The paper is ordered as follows: in section II an underlying algebraic structure of DE will be manifested, from there the schema for conceiving UFTs is going to be formulated. In section III DE will be expressed using quaternions, which allow finding simpler expressions for the usual terms in Dirac theory. Section IV will build upon section III's results and will apply section II's schema to the development of a toy theory that describes a set of particles like those in one generation of standard model with U(1) electromagnetism, SU(2) flavor and SU(3) symmetries, and a gravity-like coupling shape.

\section*{ \center II. DIRACS EQUATION ALGEBRA INSIGHTS}

In this section, an underlying algebraic structure of DE is set explicit, and it will serve as a starting point for further manipulations. This is going to be done by noting that Dirac (gamma matrices) algebra is a C$\ell_4$ Clifford algebra equivalent to $\mathfrak{su}(2)\oplus\mathfrak{su}(2)$. Therefore it can be written in terms of two $\mathfrak{su}(2)$ copies, showing an algebraic structure that relates the equation solutions to $C\ell_2$ ladder operators.

Setting conventions first, all calculations are going to be carried in natural units $\hbar=c=1$. Einstein summation is adopted unless otherwise stated, Greek indices sum over spacetime coordinates, while Latin indices run upon space.

Starting with DE,
\begin{align} \label{51}
	( i \sla \partial - m)\psi = 0,
\end{align}
which explicitly in the Weyl representation is,
\begin{align}
	 \begin{bmatrix} \begin{pmatrix} 0 & \mathbb{1} \\ \mathbb{1} & 0 \end{pmatrix} p_0 + \begin{pmatrix} 0 & \bar \sigma^j \\ \sigma^j & 0 \end{pmatrix} p_j - m\begin{pmatrix} \mathbb{1} & 0 \\ 0 & \mathbb{1} \end{pmatrix}\end{bmatrix} \psi = 0.
\end{align}
This equation can be re-expressed in terms of tensor products between two copies of Pauli matrices, a spin set of $\sigma^i$ and a chiral one of $\chi_i$, %which act over the tensor product of their corresponding groups representations, the bi-spinors,
\begin{align} \label{1}
	 \begin{bmatrix} ( \chi_1 \otimes \mathbb{1} ) p_0 - i ( \chi_2 \otimes \s^j ) p_j - m ( \mathbb{1} \otimes \mathbb{1} ) \end{bmatrix}  \psi = 0.
\end{align}

The usual approach for tackling E.q(\ref{1}) is to work column bi-spinor $\psi$, but alternatively, matrices can also be used \cite{chisholm1999, trayling2001}. Between the possibilities, are a kind of square matrices known as algebraic spinors, which have the particularity of being minimal left (right) ideals of some Clifford algebras \cite{Porteous1995}. The C$\ell_4$ algebraic spinors will emerge naturally if one decides to find solutions to the DE in terms of gamma matrices instead of bi-spinors.

Aiming to find solutions in terms of $\mathfrak{su}(2)\oplus\mathfrak{su}(2)$, a straight strategy is to solve E.q(\ref{1}) at rest frame $p_j = 0$, and then boost it. From here and on, the solutions with the column bi-spinor will be denoted by $\psi$, and the ones with algebraic elements $\Psi$. With ansatz solutions of the form 
\begin{align}  \label{2}
	\Psi=\frac{1}{2}(\mathbb{1} \pm \chi_1) e^{ \mp i m t},
\end{align}
obviating the $\otimes$ for short writing, it is clear to see that E.q(\ref{2}) is a solution for E.q(\ref{1})
\begin{align} \nonumber
	[ i \chi_1 \partial_0 - m \mathbb{1} ]  ( \chi_1 \pm \mathbb{1} ) e^{ \mp i m t} & = e^{ \mp i m t}  [ \pm \chi_1 m - m \mathbb{1} ]  ( \chi_1 \pm \mathbb{1}) 
	\\ \nonumber
	& = \pm m e^{ \mp i m t} ( \chi_1 \mp \mathbb{1} ) (\chi_1 \pm \mathbb{1}) 
	\\
	& = \mp m e^{ \mp i m t} [ (\chi_1)^2 - \mathbb{1} ] = 0.
\end{align}
Now, a generic boost can be found from the usual expression, by appealing to the isomorphism between the Weyl and the $\s_i,~\chi_i$ representations,
\begin{align}
	\Lambda_P & = \begin{pmatrix} e^{-\frac{1}{2} \s^\mu p_\mu} & 0 \\ 0 & e^{\frac{1}{2} \s^\mu p_\mu} \end{pmatrix} 
	\rightarrow \frac{1}{2}( \mathbb{1} +\chi_3) e^{-\frac{1}{2} \s^\mu p_\mu} + \frac{1}{2}( \mathbb{1} -\chi_3)  e^{\frac{1}{2} \s^\mu p_\mu},
\end{align}
with four-vectos $P=(p_0, p_1, p_2, p_3)$ and $\s^\mu = (\mathbb{1},\s^1,\s^2,\s^3)$. Furthermore the exponential factors can be worked as usual
\begin{align}
	e^{\pm\frac{1}{2} \s^\mu p_\mu} = \sqrt{ \cosh|p_0| \mp \sinh|\vec p| \hat n \cdot \vec \s } = \sqrt{ E/m \mp |\vec p|/m| } = \frac{1}{\sqrt{m}}\sqrt{p_0 \mp \vec p \cdot \vec \s},
\end{align}
and defining the chiral protectors $\chi_L=\frac{1}{2}(1+\chi_3)$ and $\chi_R=\frac{1}{2}(1-\chi_3)$, the boost matrix is given by,
\begin{align}
	\Lambda_P & = \frac{1}{\sqrt{m}} \left[ \chi_L \sqrt{p_\mu \s^\mu}  + \chi_R  \sqrt{p_\mu \bar \s^\mu} \right],
\end{align}
where $\bar \s^\mu = (\mathbb{1},-\s^1,-\s^2,-\s^3)$.
For adding the spin degrees of freedom to the solution, in analogous way to Lorentz boost the spin operator has the correspondence
\begin{align}
	S = \frac{\hbar}{2} \begin{pmatrix} \s^3 & 0 \\ 0 & \s^3 \end{pmatrix} \rightarrow \frac{\hbar}{2} \s^3,
\end{align}
with eigenvectors $s_{\up \down} = ( 1 \pm \s^3 )/2 $, since
\begin{align}
	\s^3 \frac{1}{2}( 1 \pm \s^3 ) = \pm \frac{\hbar}{2} ( 1 \pm \s^3).
\end{align}
Hence some solutions of E.q(\ref{1}) are of the form 
\begin{align} \label{3}
	\Psi & = \Lambda_P s_{\up \down}  ( \mathbb{1} \pm \chi_1) e^{ \mp i x_\mu p^\mu}.
\end{align}

What is remarkable from E.q(\ref{3}) is that $\mathcal{P}_\pm = \frac{1}{2}( \mathbb{1} \mp \chi_1)$ are actually $\mathfrak{su}(2)$ projectors, and right multiplying them by $\chi_2$, one gets
\begin{align} \label{50}
	\Psi &  = \Lambda_P s_{\up \down}  (\chi_2 \pm i\chi_3 ) e^{ \mp i x_\mu p^\mu},
\end{align}
a form of solutions that explicitly have ladders operators from the chiral algebra $\alpha_\pm = \frac{1}{2}( \mp \chi_2 + i\chi_3 )$. Now these solutions can be relate to the usual column spinors by tacking $u$'s as column eigenvectors of $\alpha_\pm$ and $\mathcal{P}_\mp$, the conventional Dirac wave functions are recovered just by left action of E.q(\ref{50}) over some $u$.

In summary, DE solutions including the ones with algebraic spinors, can be grouped as
\begin{align}\nonumber \label{4}
	\Psi^\mp & = \Lambda_P s_{\up \down} \mathcal{P}_\mp e^{ \mp i x_\mu p^\mu} 
	\\\nonumber
	\Psi_f^\mp & = \Lambda_P s_{\up \down}  \alpha_f e^{ \mp i x_\mu p^\mu} 
	\\
	\psi_f^\mp & = \Lambda_P s_{\up \down}  u_f e^{ \mp i x_\mu p^\mu},
\end{align}
taking $f$ as an arbitrary index for fermion solutions. In the column solutions the structure including projectors and ladders is absent, but it is explicit with algebraic spinor, the formers shedding light into underlying algebraic architecture. 

 The E.q(\ref{4}) expressions serve as a ground for the formulation of a unified matter field construction scheme, which consists of assuming the algebra of gamma matrices as a sub-algebra of a larger one. The logic behind this comes from E.q(\ref{4}), one can see that enlarging the $\chi_i$'s chiral Clifford algebra and hence widening the set of ladder operators, will yield a theory with more fermion fields, described simultaneously by a single Dirac like equation. The $\chi_i$'s belong to a Clifford algebra C$\ell_2$ then a natural extension is to a C$\ell_N$, however, it is not a restriction since DE will conserve its structure just by demanding that the equivalent gamma matrices follow the Dirac C$\ell_4$ algebra.

\section*{ \centering III. DOUBLE QUATERNION DIRAC EQUATION}

The present section is meant to show that changing the chiral $\chi_i$ and spin $\sigma_i$ matrices by sets of quaternions, allows casting Dirac solutions of E.q(\ref{4}) into expressions with transparent algebraic simplicity that will serve as preliminary before using octonions in the next sections.

Isomorphism between Pauli matrices and quaternions suggests a change of chiral and spin matrices by two sets of quaternions. Though it will give a space of solutions with the same dimension, they will be in terms of normed division algebras, therefore with easier to handle properties that will allow disclosing further algebraic simplicity.

Following the same idea of Rowlands nilpotent DE \cite{rowlands2003}, two idependent quaternions sets are defined
\begin{align} \nonumber \label{5}
	i \s^1 = \I ~~ ~~ i \s^2 = \J  ~~ ~~ i \s^3 = \K
	\\
	i \chi^1 = \II ~~ ~~ i \chi^2 = \JJ  ~~ ~~ i \chi^3 = \KK,
\end{align}
also two kinds of quaternions conjugates are set, for a generic element $\bb a = a_0 + a_1 \I  + a_2 \J + a_3 \K$ the spin quaternion conjugate is defined as $\bar{ \bb a} = a_0 - a_1 \I  - a_2 \J - a_3 \K$, and for an object of the type $\bb b = b_0+b_1\II+b_2\JJ +b_3 \KK$ the chiral conjugate is $ \tilde{ \bb b} =  b_0-b_1\II-b_2\JJ -b_3 \KK$, where all elements satisfy $\overline{\bb{a}\bb{a}'}=\bar{\bb{a}}'\bar{\bb{a}}$ and $\widetilde{\bb{b}\bb{b}'}=\tilde{\bb{b}}'\tilde{\bb{b}}$. Finally for completeness the hermitian of a generic object $X$, is its spin, chiral and complex conjugate, $X^\da=\tilde{{\overline X}}^*$.

Using the quaternions of E.q(\ref{5}), and defining the spacial gradient as $\bb \partial = \partial_1 \I  + \partial_2 \J + \partial_3 \K$, the DE can be written as, 
\begin{align} \label{6}
	 \begin{bmatrix} \II \partial_0 + \JJ \bb{ \partial} - m \end{bmatrix} \psi & = 0,
\end{align}
which is an equation equivalent to the one obtained by Rowlands \cite{rowlands2003}. However similarities end here since Rowlands solutions are DE in momentum and frequency domain. Here instead the E.q(\ref{4}) expression will be taken as solutions, just substituting Pauli matrices by their correspondent quaternions, where the spin factor becomes $s_{\up\down}=\frac{1}{2}(1 \pm \K)$, the projector $\mathcal{P}_\pm=\frac{1}{2}(1\pm i\II)$ and the ladder operators can be taken as $\alpha_f=\frac{1}{2}( \JJ \mp i \II) $.

For rewriting Lorentz boost, a spin four-momentum can be defined as $\bb P = p_0 - i p_1 \I - i p_2 \J - i p_3 \K $, which matches Schuricht's four-vectors shape \cite{Greiter2003} in his conplexified quaternion description of flat space-time. Then Lorentz boost of E.q(\ref{4}) can be casted as
\begin{align} \label{9}
	\Lambda_P & =  \frac{1}{2\sqrt{m}} \left[ (1-i\KK) \sqrt{ \bb P} + (1+i\KK) {\overline{ \sqrt{ \bb P}}} \right].
\end{align}

At this point quaternions pay out, by defining the left chiral projector $\chi=\frac{1}{2}(1-i\KK)$, where in E.q(\ref{9}) are recognizable the left and right Lorentz boosts $\Lambda_{L}=\chi \sqrt{ \bb P}$ and $\Lambda_{R}=\tilde{\chi} \sqrt{ \bb P}^*$,  adopting the simple form
\begin{align} \label{53}
	\Lambda_P &  = {\small \frac{1}{\sqrt{m}} } [ \chi \sqrt{ \bb P} + \tilde{\chi} \sqrt{ \bb P}^* ]
	\\
	& ={\small \frac{1}{\sqrt{m}} } \left[ \Lambda_L + \Lambda_L^*\right],
\end{align}
and the algebraic Dirac solutions of E.q(\ref{4}), transforming to each other just by
\begin{gather} 
	\Psi^+ =  \Psi^{-*} ~~~~ \Psi_f^+ = \Psi_f^{-*},
\end{gather}
breaking down complexity of expression, to clear relations in terms of conjugates, projectors and ladder operators.

\section*{ \center IV. MODIFIED DIRAC EQUATION}

This section will address an example of unified field theory, constructed from the altered DE scheme presented at the end of section II. This will be done by changing chiral quaternions of III by octonions. The properties characterization for the spawned fields lays heavily on Furey's results \cite{Furey2015, Furey2016}.

\subsection*{\center A. OCTONIONS AS CHIRAL ALGEBRA }

For an example of a unified theory constructed with section's II methodology, it is enough to choose a random algebra with a $\mathfrak{su}(2)$ sub-algebra to play the $\chi_i$'s chiral roll, enriching its inner structure by enlarging solutions' dimensions, without altering the E.q(\ref{1}) DE form. However, the simplicity of section's III expression is desirable, induced by the normed division algebra as a chiral component. Octonions are the next bigger division algebra, additionally, it has been proven that they can give account for electric and color symmetries, therefore using them for writing a DE could yields solutions that inherit those properties. Hence it is plausible to ask what if the chiral algebra is not quaternionic but octonionic. Remarkably this theory will encompass all possible division algebras $\mathbb{R}\oplus\mathbb{C}\oplus\mathbb{H}\oplus\mathbb{O}$, known as Dixon algebra. 

Octonions are not associative, a feature that is overcome in the following steps by considering only the left action of $\mathbb{C}\oplus\mathbb{O}$ on itself, since an isomorphism to $C\ell_6$ exists \cite{Furey2016}. Ergo, the choosing of octonions could be argued as merely cosmetic, since non-associativity is disregarded. Nevertheless, it is a matter of taste, rather than a semantic discussion.

\begin{figure}[h]
    \centering
    \includegraphics[width=0.4\textwidth]{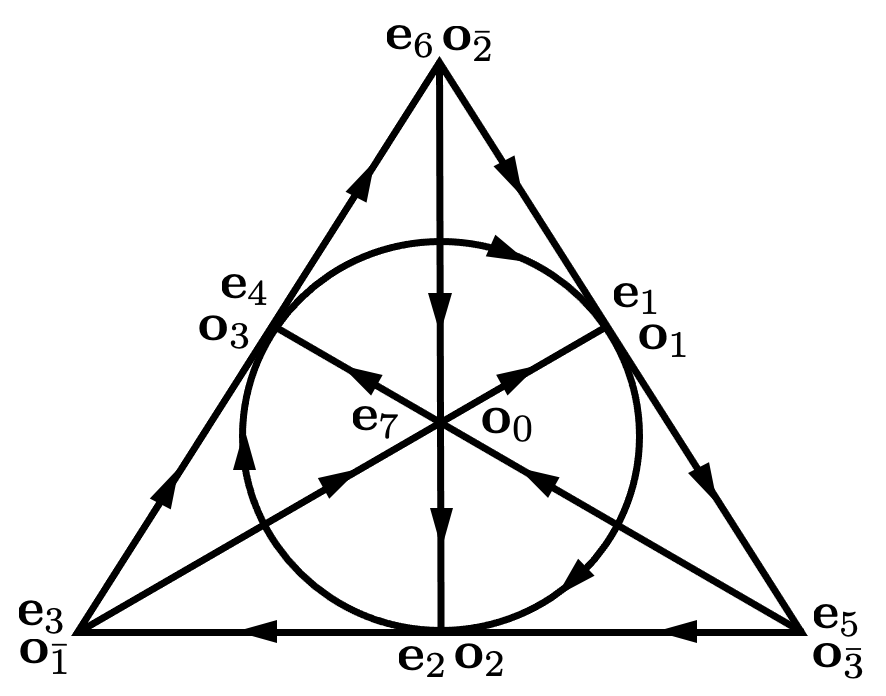}
    \caption{Octonion multiplication rules. Each trio of unities $ \mathbf{e}_{j} $ on a single segment, conforms a set of quaternions. The $\mathbf{o}_{j}$ objects are the nearest octonions unities in the diagram, under unitary transformation $U\mathbf{e}_{j}U^{-1}$.}
    \label{f1}
\end{figure}

For a brief introduction octonions have seven imaginary units $\E_j$ with multiplication rules shown in Fig.(\ref{f1}). All unities anti-conmute  $\E_j\E_i = - \E_i\E_j $, and not associative for unities belonging to different segments of Fig.(\ref{f1}), existing the possibility of $(\E_i(\E_j\E_k)) \neq ((\E_i\E_j)\E_k)$. For the simplicity, only left multiplication is going to be considered, taking as equivalents $\E_i\E_j\E_k = (\E_i(\E_j\E_k))$ \cite{dixon1994}. Therefore a generic element of octonions is $f = f_0 + \sum_{j=1}^7f_j \E_j$, and the chiral conjugate is now adopted as $\tilde{f}= f_0 - \sum_{j=1}^7f_j \E_j $. And additional to the seven unities $\E_j$, are defined the plus/minos elements as $\E_+=\E_{4}\E_{2}\E_{1}$ and $\E_-=\E_{3}\E_{6}\E_{5}$. Finally, from Fig.(\ref{f1}), it can be infer a representation, for left multiplication of two generic octonions, $f $ and $g$,
\begin{align} \label{52}
f g  =\footnotesize{
\left(\begin{array}{cccccccc}
f_0 & -f_1 & -f_2 & -f_3 & -f_4 & -f_5 & -f_6 & -f_7 \\
f_1 & f_0 & -f_4 & -f_7 & f_2 & -f_6 & f_5 & f_3 \\
f_2 & f_4 & f_0 & -f_5 & -f_1 & f_3 & -f_7 & f_6 \\
f_3 & f_7 & f_5 & f_0 & -f_6 & -f_2 & f_4 & -f_1 \\
f_4 & -f_2 & f_1 & f_6 & f_0 & -f_7 & -f_3 & f_5 \\
f_5 & f_6 & -f_3 & f_2 & f_7 & f_0 & -f_1 & -f_4 \\
f_6 & -f_5 & f_7 & -f_4 & f_3 & f_1 & f_0 & -f_2 \\
f_7 & -f_3 & -f_6 & f_1 & -f_5 & f_4 & f_2 & f_0
\end{array}\right)
\left(\begin{array}{c}
g_0  \\
g_1 \\
g_2 \\
g_3 \\
g_4 \\
g_5 \\
g_6 \\
g_7
\end{array}\right)
}.
\end{align}

Now the question is which elements to select as chiral components of E.q(\ref{3}). For such selection and clarity for further expression manipulation E.q( \ref{52})  is going to be transformed into a basis where $\E_+$ is diagonal. This is done through the unitary transformation
\begin{equation}
U = \footnotesize{
\left(\begin{array}{cccccccc}
1 & 0 & 0 & 0 & 0 & 0 & 0 & 0 \\
0 & 0 & 1 & 0 & 0 & 0 & 0 & 0 \\
0 & 0 & 0 & 0 & 1 & 0 & 0 & 0 \\
0 & 1 & 0 & 0 & 0 & 0 & 0 & 0 \\
0 & 0 & 0 & 0 & 0 & 0 & 0 & -i \\
0 & 0 & 0 & 0 & 0 & 0 & -i & 0 \\
0 & 0 & 0 & 0 & 0 & -i & 0 & 0 \\
0 & 0 & 0 & -i & 0 & 0 & 0 & 0
\end{array}\right)
},
\end{equation}
such that $\bo_\ell=U\mathbf{e}_{j}U^{-1}$ as is indicated in Fig.(\ref{f1}), yields
\begin{equation}
UfU^{-1} =
\footnotesize{
\left(
\begin{array}{cccccccc}
 f_0 & -f_2 & -f_4 & -f_1 & -i f_7 & -i f_6 & -i f_5 & -i f_3 \\
 f_2 & f_0 & -f_1 & f_4 & i f_6 & -i f_7 & i f_3 & -i f_5 \\
 f_4 & f_1 & f_0 & -f_2 & i f_5 & -i f_3 & -i f_7 & i f_6 \\
 f_1 & -f_4 & f_2 & f_0 & i f_3 & i f_5 & -i f_6 & -i f_7 \\
 -i f_7 & i f_6 & i f_5 & i f_3 & f_0 & f_2 & f_4 & f_1 \\
 -i f_6 & -i f_7 & -i f_3 & i f_5 & -f_2 & f_0 & f_1 & -f_4 \\
 -i f_5 & i f_3 & -i f_7 & -i f_6 & -f_4 & -f_1 & f_0 & f_2 \\
 -i f_3 & -i f_5 & i f_6 & -i f_7 & -f_1 & f_4 & -f_2 & f_0 \\
\end{array}
\right).
}
\end{equation}

Among this objects, it is noteworthy that $\{i\bo_0, -\bo_-,\bo_+\}$ behaves as Pauli matrices, explicitly
\begin{gather} \nonumber
	\bb o_0 = - i \footnotesize{\left(
\begin{array}{cccc}
 0 & 0 & \mathbb{1} & 0 \\
 0 & 0 & 0 & \mathbb{1} \\
 \mathbb{1} & 0 & 0 & 0 \\
 0 & \mathbb{1} & 0 & 0 \\
\end{array}
\right)}
\\ \label{58}
	\bb o_+ = \footnotesize{\left(
\begin{array}{cccc}
 \mathbb{1}  & 0 & 0 & 0 \\
 0 & \mathbb{1}  & 0 & 0 \\
 0 & 0 & -\mathbb{1}  & 0 \\
 0 & 0 & 0 & -\mathbb{1}  \\
\end{array}
\right)}
~~~~~~~~~
	\bb o_- = \footnotesize{  i \left(
\begin{array}{cccc}
 0 & 0 & \mathbb{1} & 0 \\
 0 & 0 & 0 & \mathbb{1} \\
 -\mathbb{1} & 0 & 0 & 0 \\
 0 & -\mathbb{1} & 0 & 0 \\
\end{array}
\right)},
\end{gather}
which gives a straightforward identification for the chiral matrices
\begin{gather} \label{8}
	\chi_1 \rightarrow i\bo_0 ~~~~~ \chi_2 \rightarrow - \bo_- ~~~~~ \chi_3 \rightarrow \bo_+.
\end{gather}
From E.q(\ref{8}), the E.q(\ref{6}) DE, with octonions (chiral), and quaternions (spin) becomes in 
\begin{align} \label{13}
	 \begin{bmatrix} \bo_0 \partial_0 + i \bo_- \bb \partial + m \end{bmatrix} \psi = 0.
\end{align}
This form yields the direct identification for gamma matrices,
\begin{align}
	\g^0 = i \bo_0 ~~~~~ \g^1= \bo_- \I ~~~~~ \g^2= \bo_- \J ~~~~~ \g^3 = \bo_- \K ~~~~~ \g^5 = \bo_+ ,
\end{align}
in consequence permits to write Dirac Lagrangian in its usual form with a thick-bar notation for adjoint spinors $ \thickbar\psi = \psi^\da \g^0$, as
\begin{align} \label{12}
	  \mathcal{L} =  \thickbar \psi \left[ i\g^\mu \partial_\mu - m \right] \psi = 0.
\end{align}
It is worth highlighting that the Lagrangian conserves its form from original Dirac theory, which make the treatment analogous to the unmodified one. With that said, it is expected that discrete symmetries of original theory hold. Now, even though the form is alike, the $\g^\mu$ are $16\times16$ complex matrices ($\mathbb{C}\oplus\mathbb{H}\oplus\mathbb{O}$), providing the theory with a richer structure for internal degrees of freedom, hence for describing more kinds of fermions.

\subsection*{\center B. SOLUTIONS }

With the Lagrangian set, the next step is finding the fields it describes. Starting from the projector solution $\Psi^\mp = \Lambda s_{\up\down} \mathcal{P}^\mp  e^{ \mp i x_\mu p^\mu }$, with $\mathcal{P}_\mp = \frac{1}{2}( 1 \mp i  \bb o_0 ) $. Spin component remains unchanged, and left chiral projector $\chi=\frac{1}{2}(1-\bo_+)$, E.q(\ref{53}) expression holds, enabling to write the Lorentz boost,
\begin{align} \label{59}
	\Lambda_P &  =  \frac{1}{\sqrt{m}} \left[ \chi \sqrt{ \bb P} + \widetilde{(\chi \sqrt{ \bb P})^*}\right].
\end{align}
The following stage is to find the $\mathcal{P}_\mp$ algebraic spinors eigenvectors, the ladders operators or arches are defined for this
\begin{align} \label{58}
	\alpha_{\ell}^\da = \frac{1}{2}( - \bb o_\ell + i \bb o_{\bar \ell } ) ~~~~~ \alpha_{\ell} = \frac{1}{2}( \bb o_\ell + i \bb o_{\bar \ell} ),
\end{align}
given a $C\ell_6$ fermion algebra,
\begin{align}\label{10}
	\{  \alpha_{\ell}^\da ,  \alpha_{j} \} = \delta_{\ell j}.
\end{align}
An orthogonal set of eigenvectors for the projector $\mathcal{P}^\pm$ from these operators can be constructed. This can be done first by realizing that eigenvectors are the nilpotent operators
\begin{align}
	\omega^\da = \alpha_1^\da \alpha_2^\da \alpha_3^\da ~~~~~~~~ \omega = \alpha_3 \alpha_2 \alpha_1.
\end{align}
Upon them other objects can be built via successive application of E.q(\ref{10}) ladder operators over the idempotent $\omega\omega^\da$, this leads to linear combinations of algebraic spinors (ideals) \cite{Furey2016}
\begin{gather}
	\begin{array}{c c} \label{55}
	\Psi = & \ell_1 \omega\omega^\da 
	\\ 
	&+ \bar q^r_1 \alpha_1^\da \omega\omega^\da + \bar q^g_1  \alpha_2^\da \omega\omega^\da + \bar q^b_1 \alpha_3^\da \omega\omega^\da
	\\
	&+q^r_2 \alpha_3^\da  \alpha_2^\da \omega\omega^\da +q^g_2  \alpha_1^\da  \alpha_3^\da \omega\omega^\da +q^b \alpha_2^\da  \alpha_1^\da \omega\omega^\da
	\\
	& + \bar \ell_2 \alpha_3^\da \alpha_2^\da  \alpha_1^\da \omega\omega^\da,
	\end{array}
\end{gather}
where, for coming reasons, $\ell$ is for leptons, $q$ for quarks, and the small over-bar is for antiparticles. For constructing column spinorial solutions it is enough to find the eigenvector $u$ of $\omega\omega^\da$, with eigenvalue different to zero, and from it finding the other ones by analogue construction to the algebraic spinor case. The corresponding eigenvector is $u = (1, 0, 0, 0, 1, 0, 0, 0)^\top,$ hence by successive application of the ladder operators,
\begin{equation} \label{11}
\begin{array}{c}
\big( \ell_1 + \bar q_1^r \alpha_1^\da + \bar q_1^g \alpha_2^\da + \bar q_1^b \alpha_3^\da
\\
~~~~~~~~~~~ + \alpha_2^\da  \alpha_3^\da q_2^r  + \alpha_3^\da  \alpha_1^\da  q_2^g + \alpha_1^\da  \alpha_2^\da q_2^b  +\bar \ell_2 \alpha_3^\da \alpha_2^\da \alpha_1^\da \big) u = 
\end{array}
\footnotesize{
\left(\begin{array}{c}
\ell_1+\bar \ell_2  \\
q_2^g+\bar q_1^g \\
q_2^b+\bar q_1^b \\
q_2^r+\bar q_1^r  \\
\ell_1-\bar \ell_2 \\
q_2^g-\bar q_1^g \\
q_2^b-\bar q_1^b  \\
q_2^r-\bar q_1^r 
\end{array}\right)
}.
\end{equation}

From E.q(\ref{11}) and E.q(\ref{4}), a generic column solution can be written as $\psi = (e^{\pm ip_\mu x^\mu}\Lambda_P s_{\up\down})(\alpha u) $ where $\alpha$ is any product of $\alpha_i$'s, letting to classified the eight solutions, into four kinds
\begin{align} \nonumber 
	\ell_1 & = (e^{ - i p_\mu  x^\mu} \Lambda_P s_{\up\down} )( \omega \omega^\da  u )
	\\ \nonumber
	\bar \ell_2 & = (e^{ i p_\mu x^\mu} \Lambda_P s_{\up\down} )( \alpha_3^\da \alpha_2^\da  \alpha_1^\da \omega \omega^\da u )
	\\ \nonumber
	\bar q_1^c  & = (e^{ i p_\mu  x^\mu} \Lambda_P s_{\up\down} )( \alpha_i^\da \omega \omega^\da u)
	\\ \label{22}
	q_2^c  & = (e^{ - i p_\mu x^\mu} \Lambda_P s_{\up\down} )( \alpha_i^\da  \alpha_j^\da \omega \omega^\da u ),
\end{align}
showing an evident structure, with separable spacetime/spin components, from the $\alpha_j$'s.

As it is shown in subsection C, each E.q(\ref{22}) solution type describes a particle class, where the main distinguishable feature is the $\alpha_j$'s factors, endowing them with charges and a sort of flavor notion, from there the suitable label of arches. Then the theory solutions meets a speculated correspondence of ideals$\sim$particles, $\mathbb{C}\oplus\mathbb{H} \sim$Lorentz and $\mathbb{C}\oplus\mathbb{O} \sim$internal degrees of freedom \cite{Furey2016}.

\subsection*{\center C. SOLUTION CHARACTERIZATION AND SYMMETRIES}

The next step is to identify each solution with a kind of particle, for such task charge operators can be formulated from the arches.

An $U(1)$ generator that behaves as an electric charge can be constructed as
\begin{equation}
	Q^u = \frac{1}{3} \sum_{i} \alpha_{i}^{\dagger} \alpha_{i},
\end{equation}
$SU(3)$ generators are given by,
\begin{equation}
	\begin{array}{ll}
		\Lambda_{1}^u=-\alpha_{2}^{\dagger} \alpha_{1}-\alpha_{1}^{\dagger} \alpha_{2} & \Lambda_{2}^u=i \alpha_{2}^{\dagger} \alpha_{1}-i \alpha_{1}^{\dagger} \alpha_{2} \\
		\Lambda_{3}^u=\alpha_{2}^{\dagger} \alpha_{2}-\alpha_{1}^{\dagger} \alpha_{1} & \Lambda_{4}^u=-\alpha_{1}^{\dagger} \alpha_{3}-\alpha_{3}^{\dagger} \alpha_{1} \\
		\Lambda_{5}^u=-i \alpha_{1}^{\dagger} \alpha_{3}+i \alpha_{3}^{\dagger} \alpha_{1} & \Lambda_{6}^u=-\alpha_{3}^{\dagger} \alpha_{2}-\alpha_{2}^{\dagger} \alpha_{3} \\
		\Lambda_{7}^u=i \alpha_{3}^{\dagger} \alpha_{2}-i \alpha_{2}^{\dagger} \alpha_{3} & \Lambda_{8}^u=-\frac{1}{\sqrt{3}}\left[\alpha_{1}^{\dagger} \alpha_{1}+\alpha_{2}^{\dagger}\alpha_{2}-2 \alpha_{3}^{\dagger} \alpha_{3}\right],
	\end{array}
\end{equation}
and the complex conjugate results in other set of independent generators,
\begin{equation}
	Q^d =  Q^{u*} = \frac{1}{3} \sum_{i} \alpha_{i} \alpha_{i}^{\dagger} ~~~~~~~~ \Lambda_{i}^d = \Lambda_{i}^{u*}.
\end{equation}
The confluence of all those quadratic terms, is the general linear combination:
\begin{equation}
\sum_{\mathcal{H}} \mathcal{H}= a^u Q^u+\sum_{i=1}^{8} g_{i}^u \Lambda_{i}^u + a^d Q^d+\sum_{i=1}^{8} g_{i}^d \Lambda_{i}^d,
\end{equation}
with $a_j^{u/d}$'s and $g_i^{u/d}$'s some $\mathbb{R}$ coefficients.

The algebraic spinors of E.q(\ref{55}) under the action of the above generators, behaves as representations of standard model particles, assigning the ID of electron $\mathcal{E}$, neutrino $\mathcal{V}$, antiquark up $\bar{ \mathcal{U}}$, antidown $\bar{\mathcal{D}}$, and so on \cite{Furey2015}. However, they are identified as up $S^u$ or down $S^d$ weak-isospin particles, depending on which generators act on them as it is consigned in Tab.(\ref{t1}). As can be seen, electrons and quarks are excitations from what can be identified as a neutrino, which plays as a sort of vacuum \cite{Furey2015}.
\begin{table}[h]
\centering
\begin{tabular}{ ccc|ccc|c }
$ \underline{Q}^u$ & $\underline{\Lambda}^u $ & $\underline{S^u~ID}$ & $-\underline{Q}^d$ & $-\underline{\Lambda}^d$ & $\underline{S^d~ID}$ & $\underline{\alpha} $ \\
$ 0$ & $1$ & $\mathcal{V}$ & $-1$ & $1$ & $\mathcal{E}$ &$ \omega \omega^{\dagger} $ \\
$ 1 / 3$ & $\bar{3}$ & $\bar{\mathcal{D}}$ & $-2 / 3$ & $\bar{3}$ & $\bar{\mathcal{U}}$ & $\alpha_{i}^{\dagger} \omega \omega^{\dagger} $ \\
$ 2 / 3$ & $3$ & $ \mathcal{U}$ & $-1 / 3$ & $3$ & $ \mathcal{D}$ & $\alpha_{i}^{\dagger}  \alpha_{j}^{\dagger} \omega \omega^{\dagger} $ \\
$ 1$ & $\bar1$ & $ \mathcal{E}^+$ & $0$ & $\bar1$ & $ \bar{\mathcal{V}}$ & $\alpha_{i}^{\dagger} \alpha_{j}^{\dagger} \alpha_{k}^{\dagger} \omega \omega^{\dagger} $
\end{tabular} 
\captionsetup{justification=centering}
\caption{\label{t1} Particle properties under generators action. $Q^{u/d}$ is electric charge operator for isospin up/down. For $\Lambda$ generators $\bar3$, means that the corresponding $\alpha$ transform as color anti-triplet, $1$ for color singlet, et cetera.}
\end{table}

It will be clear later how to include both (weak-)isospins subspaces in a single description, for now it is illustrative enough to go ahead with up isospin generators. For incorporating interactions one can start with 
\begin{equation}
	\sum_{\mathcal{H}^u }\mathcal{H}^u= a Q^u+\sum_{i=1}^{8} g^i \Lambda_{i}^u.
\end{equation}
From here a gauge covariant derivative can be constructed, making the equation invariant under local phase transformation in the usual Yang-Mills form. Promoting position dependency for $a=a(x)$ and $g=g_i(x)$, for a transformation
\begin{align}
	 \Lambda_P s_{\up\down} \alpha e^{\pm ip_\mu x^\mu} \rightarrow \Lambda_P s_{\up\down} \left( e^{-i \mathcal{H}^u} \right)\alpha e^{\pm ip_\mu x^\mu}.
\end{align}
 DE is not invariant under such transformation,
\begin{align}
	&  \left[ i\g^\mu \partial_\mu - m \right]  \Lambda s_{\up\down} \left( e^{i \mathcal{H}^u} \right)\alpha e^{\pm ip_\mu x^\mu} = 0
	\\
	& \left[ i\g^\mu \left( \partial_\mu + i \{ \Lambda \partial_\mu\mathcal{H}^u \Lambda^{-1} \} \right) - m \right] \Lambda s_{\up\down}\alpha e^{\pm ip_\mu x^\mu} \neq  0,
\end{align}
to overcome this, is added the counterterm,
\begin{align}
	\partial_\mu\mathcal{H}^u = A_\mu^u Q^u + \sum_{i=1}^{8} G_\mu^u \Lambda_{i}^u,
\end{align}
with $A_\mu=\partial_\mu a$ and $G_\mu^i = \partial_\mu g^i$. Hence the covariant derivative acting on the particles with isospin up should be defined like
\begin{align}
	D_\mu^u =\partial_\mu - i \Lambda_P \left( A_\mu^u Q^u + \sum_{i=1}^{8} G^u \Lambda_{i}^u \right) \Lambda_P^{-1},
\end{align}
where the $\Lambda_P$ has been added, regarding that $Q^u$ and $\Lambda_{i}^u$ are now matrices which do not commute with Lorentz boost.

Now E.q(\ref{12}) produces solutions with same masses, to give a different mass to each kind of particle, an ad hoc isospin up mass matrix $M^u$ can be constructed as
\begin{gather}  \label{15}
	\begin{array}{c c}
	M^u  = & 
	m_\mathcal{V} \omega\omega^\da 
	\\
	&+ m_{\mathcal{D}} (\alpha_1^\da \omega\omega^\da \alpha_1 +  \alpha_2^\da \omega\omega^\da \alpha_2 + \alpha_3^\da \omega\omega^\da \alpha_3 )
	\\
	&+ m_{\mathcal{U}} (\alpha_3^\da  \alpha_2^\da \omega\omega^\da  \alpha_3  \alpha_2 + \alpha_1^\da  \alpha_3^\da \omega\omega^\da  \alpha_1 \alpha_3 + \alpha_2^\da  \alpha_1^\da \omega\omega^\da \alpha_2 \alpha_1 )
	\\
	&+ m_\mathcal{E} \omega^\da \omega,
	\end{array}
\end{gather}
and also the $1/\sqrt{M^u}$ matrix, as $M^u$ with all $m_f\rightarrow m_f^{-1/2}$. In a similar way for $Q^u$ and $\Lambda_i^u$ generators, the downs isospin analogue is given by the complex conjugate $M^d = M^{u*}$ and $1/\sqrt{M^d} = 1/\sqrt{M^{u}}^*$. Introduction of these elements could be seen as enlarging the algebra, if products between algebra elements and $M$ are taking into account, however this will not be explored in this paper.

Putting together \Eq{13}, the gauge covariant derivative \Eq{14} and the mass matrix \Eq{15}, arrive to an equation that contains the particle spectrum of isospin up subspace of one standard model generation, and the corresponding equation for isospin down is gotten by changing $u \rightarrow d$. In summary
\begin{align} \label{16}
	\left[ i\g^\mu D_\mu^{u/d} - M^{u/d} \right] \psi^{u/d} = 0,
\end{align}
and the solutions only change by $1/\sqrt{m}\rightarrow1/\sqrt{M^{u/d}}$ in \Eq{59} Lorentz boost. \Eq{16} describes a unified matter field theory with electromagnetic and strong like forces, but just for a half of one standard model particle generation isospin up or down at a time.

Additionally, it is pointed out that it is possible to write the DE in terms of electric charge operator, considering
\begin{align}
	  \bo_0 =  \bo_3  \bo_2  \bo_1  \bo_{\bar1}  \bo_{\bar2}  \bo_{\bar3},
\end{align}
and using $ \bo_j = \alpha_j -  \alpha_ j^\da $, $ i \bo_{\bar j} = \alpha_j^\da +  \alpha_ j $, it is possible to express $ \g^0$ in terms of the electric charge operator $Q^u$
\begin{align}
	  \g^0 = \left( - 1 + 2 \left( 10 Q^u - 9 Q^{u2} + 2 Q^{u3} \right) \right),
\end{align}
which quaintly looks like first orders of an expansion.

\subsection*{ \center D. ONE GENERATION UNIFIED FIELD LAGRANGIAN}
Although the octonionic DE presented in \Eq{16} succeeded in reproducing particles that can be identified as two leptons and two quarks under electromagnetic and strong force, it describes only a isospin up or down subspace of solutions, meaning that they only have a particle or antiparticle of each type. A clue for a possible way of overcoming this is CPT symmetry, which should transform \Eq{16} between particles up and down isospin versions exchanging particles and antiparticles.

From the usual CPT transformation it is given the form
\begin{equation}\label{17}
	\psi(t,\vec x) \overset{\mathrm{CPT}}{\rightarrow} i \g^5 \psi(-t,-\vec x),
\end{equation}	
but in the normal DE the $Q$ and $\Lambda_i$ generators commute with $\g^5 = \bo_+$, in this case instead, the formers belong to the same algebra having
\begin{align}\label{21}
	Q^u \g^5 = \g^5 Q^d ~~~~ \Lambda^u \g^5 = \g^5 \Lambda^d ~~~~ M^u \g^5 =  \g^5 M^d.
\end{align}
Taking into account that the CPT transformation also transform between isospin operators, the correspondence between particles and antiparticles remains as
\begin{align} \label{56}
	\left[ i\g^\mu D_\mu^{u} - M^{u} \right] \psi^{u} \overset{\mathrm{CPT}}{\leftrightarrow} \left[ i\g^\mu D_\mu^{d} - M^{d} \right] \psi^{d}.
\end{align}
Then an invariant Lagrangian (density) under CPT transformation, can be formulated containing both isospin version of \Eq{16},
\begin{align} \label{20}
	  \mathcal{L} = & \thickbar \psi^u \left[ i\g^\mu D_\mu^u  - M^u  \right] \psi^u +  \thickbar \psi^d \left[ i\g^\mu D_\mu^d - M^d \right] \psi^d,
\end{align}

For short writing from here and on it is going to be suppressed the super-index $u$ on all up isospin operators, and Dirac operator is going to be denoted by
\begin{align} \label{57}
	\mathcal{D} = - i\g^\mu D_\mu  + M = \bo_0 D_0 + i \bo_- \bb D + M.
\end{align}
The \Eq{57} Lagrangian can be reformulated in a simpler form. It is point out that for this from \Eq{58} for any arche $\alpha_j^*= -\alpha_j^\da$. This is the reason why for all generators in gauge derivatives and mass matrix the conjugate complex transform between isospins subspaces. Also, applied to the solutions of \Eq{22}, complex conjugate relates wave functions in both isospin subspaces. However, \Eq{59} Lorentz boosts are not invariant under complex conjugate, this is fixed noting that they are invariant under simultaneous complex and quaternion conjugate, with quaternion conjugate not affecting the arches. From here it is straightforward to write \Eq{56} into the succinct formula
\begin{align}  \label{14}
	  \mathcal{L} = & \thickbar \psi \mathcal{D} \psi +  \overline{\left( \thickbar \psi \mathcal{D} \psi \right)^*}.
\end{align}
In this expression the relation between isospin sub-spaces is explicit. But as it is known Lagrangians are not unique, a further form can be ideated which allows an additional interpretation. Summing 0 terms as Dirac operator powers
\begin{align} 
	 \sum_{n=1}^\infty \frac{(i \mathcal{D})^n}{n!} \psi = & 0 ,
\end{align}
equivalent to
\begin{align} \label{60}
	 e^{i\mathcal{D}} \psi = & \psi,
\end{align}
and it corresponds to the alternative Lagrangian
\begin{align}  \label{64}
	\mathcal{L} & = \thickbar{\psi} e^{i\mathcal{D}} \psi - \thickbar{\psi} \psi  + q.c.c,
\end{align}
where $q.c.c$ is for quaternion-complex conjugate. This Lagrangian describes a unified matter field theory with endogenous gauge groups $U(1)$ electromagnetism and $SU(3)$ color, alike to a full generation of standard model particles, and CPT symmetry is equivalent to a quaternion-complex conjugation. It is noteworthy, that the matter fields are the ones unified while the interactions arise from the symmetry properties of the fermion fields, rather than the imposition of a gauge master group. Additionally, the form of equation \Eq{60} and \Eq{64} Lagrangian exhibits the feature that the system symmetries and dynamics emanate from demanding invariance under phase transformation $\psi \rightarrow \psi'=e^{i\mathcal{D}}\psi$. This will come in handy when introducing flavor interactions.

\subsection*{ \center E. ON FLAVOR INTERACTION AND FRAME FIELDS}

For the theory developed so far, is natural to ask if it can be extended to include a flavor changing interaction by using elements of the algebra. A path for this is insinuated by \Eq{20} being conveniently arranged separating isospin subspaces, which suggest to group $\psi^u$ and $\psi^d$ in an isospin doublet, 
\begin{align} \label{45}
	   \mathcal{L}  =
	   \footnotesize{
	   \begin{pmatrix} \thickbar{\psi} & \overline{\thickbar{\psi}^*} \end{pmatrix} \begin{pmatrix} \mathcal{D} & 0 \\  0 & \overline{\mathcal{D}^*}  &  \end{pmatrix}  \begin{pmatrix} \psi \\ \overline{\psi^*} \end{pmatrix}
	   },
\end{align}
at first glance, it seems that for a SU(2) flavor change elements outside of the algebra are needed, though this can be achieved with elements inside it as will be subsequently proven. A hint is given by $\g^5= \bo_+$ in the CPT transformation which connects isospin doublets, allowing the speculation of a relation between flavor force and space-time.

Exploring the action of \Eq{58} chiral matrices $i\bo_0, -\bo_-,\bo_+$ on algebraic solutions of \Eq{55}, it is identifiable that they act as $\mathfrak{su}(2)$ flavor changing matrices
\begin{align} \label{31}
	\footnotesize{ \tau_1 = -\bo_+ =  -\g^5 \equiv \begin{pmatrix} 0 & 1 \\ 1 & 0 \end{pmatrix} ~~~  \tau_2 = \bo_- = i \g^5 \g^0  \equiv \begin{pmatrix} 0 & -i \\ i & 0 \end{pmatrix} ~~~ \tau_3 =  i \bo_0 = \g^0  \equiv \begin{pmatrix} 1 & 0 \\ 0 & -1 \end{pmatrix} },
\end{align}
on doublets,
\begin{align} \label{46}
	\begin{pmatrix} \ell_1 \\ \ell_2 \end{pmatrix} ~,~ \begin{pmatrix} q_2 \\ q_1 \end{pmatrix}.
\end{align}
In contrast to the standard model left weak force, here the generators act on left and right particles equally, and they are also in terms of gamma matrices. An attempt to add left chiral projectors to mimic the usual weak force is reasonable, however the $\chi \tau_i$'s do not form a $\mathfrak{su}(2)$ algebra neither over algebraic or column spinors. 

From \Eq{31} and \Eq{31} the tentative next step is to use Yang-Mills scheme, for constructing an invariant theory under the $\tau_i$'s induced SU(2) transformation
\begin{equation}
	\begin{pmatrix} \psi \\ \overline{{\psi}^*} \end{pmatrix} \rightarrow e^{\frac{i}{2}\text{w}_i\tau_i} \begin{pmatrix} \psi \\ \overline{{\psi}^*} \end{pmatrix}.
\end{equation}
However, this will not succeed since the Lagrangian is not globally invariant under the transformation due to $\g^\mu$'s and $\tau_i$'s do not commute being one and the same from \Eq{31}.

Now if one insists on exploring the inclusion of an interaction with\Eq{58} chiral generators, the Lagrangian can be made invariant under certain conditions, which produces curious expressions to be derived.

To find how a flavor changing interaction can be introduced into the theory, one can insert a new $Y$ matrix to discompose the mass matrix $ - M = \text{b} Y + \text{w}_1 \tau_1 $, with b,$\text{w}_1\in\mathbb{R}$. Regarding the equivalence given in \Eq{31} Dirac operator can be recast as \begin{align}
	\mathcal{D} = - i\tau_3 D_0 + i \tau_2 \bb D - \text{w}_1 \tau_1 - \text{b} Y,
\end{align}

in conjunction with \Eq{60}
\begin{align} 
	 e^{-i( \text{w}_1 \tau_1 - i \tau_2 \bb D + i\tau_3 D_0  + \text{b} Y )} \psi = &  \psi.
\end{align}
A form that tempts the unconventional eigenvalues associations of $ i D_0 $ and $ i \bb D $, to $E \equiv \text{w}_3$ and $- \bb P  \equiv \text{w}_2$ respectively. They lead to see DE as an spinor invariance down flavor phase transformations
\begin{align} \label{61}
	e^{-i(  \text{b} Y + \vec \tau \cdot \vec{\text{w}} )} \psi = & \psi,
\end{align}
constrained by the equivalent Einstein energy-momentum relation 
\begin{equation} \label{62}
	\text{w}_3 = \pm \sqrt{|\text{w}_2|^2+ ( \text{w}_1 +\text{b} )^2}.
\end{equation}
Just by the form of \Eq{62} $Y$ could be identified as a sort of (weak) hypercharge, and its relation to mass $M = \text{w}_1 \tau_1 - \text{b} Y $ looks more or less like a Gell-Mann–Nishijima formula. However, its nature and relation to the already ad hoc mass matrix remain unexplored. On the other hand, $Y_\text{w}$ hypercharge can be defined by the usual Gell-Mann–Nishijima relation equivalent to $Y_\text{w} = \tau_3 - 2 Q $, nonetheless $Y_\text{w}$ values for right-handed particles will not correspond to those of standard model since $\tau_i$'s acts on left and right indistinguishably.

At this point, flavor force can be included by gauge covariant derivatives, making the $\text{w}_i$'s phases spacetime functions 
\begin{equation}\label{49}
	\sum_{\mathcal{H}} \mathcal{H}= Q a(x) + \vec \tau \cdot \vec{\text{w}}(x) + \sum_{i=1}^{8} \Lambda_{i} g^{i}_\mu(x),
\end{equation}
changing the gauge covariant derivative to
\begin{equation} \label{63}
	D_\mu = \partial_\mu - i \Lambda_P \left( A_\mu Q + i \vec \tau \cdot \vec W_\mu + \sum_{i=1}^{8} G \Lambda_{i} \right) \Lambda_P^{-1},
\end{equation}
with $ \vec W_\mu=\partial_\mu \vec{\text{w}}$, meeting \Eq{62} constrain. Promoting the flavor phases from global to local, carries the abnormal consequence of energy and momentum induced spacetime dependency since $\text{w}_3(x) \equiv E(x) $ and $  \text{w}_2(x) \equiv - \bb P(x)$, explicitly yielding into DE
 \begin{equation} \label{65}
	(i\g^0E(x)+\g^j p_j(x)-M)\psi=0.
\end{equation}

Now the question is what the spacetime dependency could mean. A possible interpretation came from general relativity.  Gravity-like effects are introduced into DE, by the hand of $e_{a}^{\mu}(x)$ (vierbeins) frame fields, which are directly coupled to gamma matrices, giving the curve spacetime DE
 \begin{equation} \label{67}
	[i(\g^a e_{a}^{0})D_0+i(\g^a e_{a}^{j})D_j-M]\psi=0,
\end{equation}
where $D_\mu$ is a covariant derivative including also spin connections, and Latin indices of $e_{a}^{\mu}$, are for flat Lorentzian spacetime, while Greek are for curve relating Minkowski metric $\eta^{ab}$ to a curve metric $g^{\mu\nu}$ by $g^{\mu \nu}(x)=e_{a}^{\mu}(x) e_{b}^{\nu}(x) \eta^{a b}$. Reordering \Eq{67}
 \begin{equation} \label{66}
	[i\g^0 (e_{0}^{\mu}D_\mu)+i\g^j (e_{j}^{\mu}D_\mu)-M]\psi=0.
\end{equation}
Comparing term by term between \Eq{65} and \Eq{66}, suggests the straight forward correspondence of $E(x)$ and $p_j(x)$, as eigenvalues of $e_{0}^{\mu}D_\mu$ and $e_{j}^{\mu}D_\mu$ respectively, setting in this theory a direct relation between flavor bosons and framefields. Further studies in how gravity could behave are needed.\section*{ \center V. CONCLUSIONS}

By breaking down the Dirac equation in terms of chiral and spin $\mathfrak{su}(2)$ matrices, it was possible to express its solutions using projectors and ladder operators. A construction scheme for formulating unified field theories was proposed from the obtained forms. This consists in taking an algebra where Dirac gamma matrices are sub-algebra elements, done by enlarging the chiral matrices, introducing new degrees of freedom that allow the resultant Dirac-like equation to describes several fermions in a unified matter field. The fields could have symmetries endowed by inheritance of the algebra, and forces are introduced by Yang-Mills covariant derivatives.

As an illustrative example, a unified matter field theory was constructed using quaternions as spin matrices and left multiply octonions as chiral components. These elections were motivated by the fact that they are normed division algebras, therefore have nice properties to work out, yielding concise expression by use of conjugates, projectors, and ladder operators. The resulting theory describes a full generation of standard model particles, under U(1) electromagnetism, SU(2) flavor and SU(3) color gauge groups, and remarkably SU(2) flavor enters with abnormal behavior, coupling its bosons to gamma matrices as frame fields, opening connections to gravity. It is noteworthy that what is unified are the fermion fields, and forces emerge from inherent symmetries, rather than from a master gauge group.

Demanding Dirac equation to describe particles with different masses, an ad hoc mass matrix had to be included, which later is shown to be equivalent to introducing a matrix that looks like a weak hypercharge analog. However, the model does not give account for electroweak force, and the Gell-Mann–Nishijima formula does not yield the correct weak isospin charges.

Based on what has been developed through this paper future paths of investigation could be exploration of the dynamics induced by frame fields, algebra enlargement by mass ($Y$) matrix, and further algebras for the proposed scheme to emulate standard model particles more accurately.

\subsection*{\center ACKNOWLEDGMENTS }
The author is deeply indebted to P. Arevalo for correcting the manuscript and would like to thank W. E. Salazar for valuable discussions and feedback.

\bibliographystyle{ieeetr}
\bibliography{refs}

\end{document}